# Semi-blind source separation using convolutive transfer function for nonlinear acoustic echo cancellation

Guoliang Cheng,[1,2] Lele Liao,[1,2] Kai Chen,[1] Yuxiang Hu,[2] Changbao Zhu,[2] and Jing Lu[1,2,a]

[1] Key Laboratory of Modern Acoustics, Institute of Acoustics, Nanjing University, Nanjing 210093, China.

[2] NJU-Horizon Intelligent Audio Lab, Horizon Robotics, Nanjing 210038, China.

Running title: SBSS using CTF for NAEC

**Abstract**: The recently proposed semi-blind source separation (SBSS) method for nonlinear acoustic echo cancellation (NAEC) outperforms adaptive NAEC in attenuating the nonlinear acoustic echo. However, the multiplicative transfer function (MTF) approximation makes it unsuitable for real-time applications especially in highly reverberant environments, and the natural gradient makes it hard to balance well between fast convergence speed and stability. In this paper, we propose two more effective SBSS methods based on auxiliary-function-based independent vector analysis (AuxIVA) and independent low-rank matrix analysis (ILRMA). The convolutive transfer function (CTF) approximation is used instead of MTF so that a long impulse response can be modeled with a short latency. The optimization schemes used in AuxIVA and ILRMA are carefully regularized according to the constrained demixing matrix of NAEC. Experimental results validate significantly better echo cancellation performance of the proposed methods.

[a] Electronic mail: lujing@nju.edu.cn

## I. INTRODUCTION

Acoustic echo cancellation (AEC) plays an important role in speech front-end processing. The usual linear AEC methods [1,2] perform well on echo with pure linear acoustic transfer function. However, the nonlinear distortion in many practical applications is not negligible, causing significant deterioration of the linear AEC [3,4]. Various nonlinear AEC (NAEC) methods have been proposed based on Volterra filters [5,6], function link adaptive filters [7], particle filters [8,9], state-space frequency-domain adaptive filters [10–12] and kernelized adaptive filters [13]. These methods are designed by combining the adaptive filter with a pre-assumed nonlinear model, which often mismatches the actual nonlinear model and thus challenges the behavior of the adaptive filter.

Recently, we proposed an NAEC method based on semi-blind source separation (SBSS) [14]. It roots from independent vector analysis (IVA) [15,16], which can better solve the permutation problem than independent component analysis (ICA)-based methods and has been proven to be one of the most effective blind source separation (BSS) methods. Unlike the method based on adaptive filters, which aims at identifying the nonlinear transfer function directly, the SBSS method is based on the independence assumption between the reference signal and the near-end signal, and therefore less sensitive to the model mismatch. However, the method proposed in Ref. 14 has two drawbacks. First, the SBSS for NAEC is derived in the short-time Fourier transform (STFT) domain based on the multiplicative transfer function (MTF) approximation [17], which relies on the assumption of a long STFT analysis window that can cover the length of the system impulse response. However, as the long analysis window leads to a long latency, the MTF-based SBSS is not suitable for real-time NAEC, especially in highly reverberant environments. Second, it is hard to determine a proper step size for the adopted natural gradient algorithm to balance well between the convergence speed and the stability. In BSS field, auxiliary-function-based IVA (AuxIVA) [18–

[21] and independent low-rank matrix analysis (ILRMA) [22–24] have been proven to have better performance than IVA based on natural gradient due to their iterative updating process without explicit step sizes and more flexible source models. However, their optimization schemes cannot be used in SBSS directly due to the special structure of the demixing matrix.

In this paper, we first improve the SBSS using the convolutive transfer function (CTF) approximation [25, 26], which can efficiently model long impulse responses using short time frames. Then we carefully regularize the optimization schemes used in AuxIVA and ILRMA based on the constrained structure of the demixing matrix and combine them into the SBSS method, resulting in better convergence behavior. Both methods are designed in online form to facilitate their real-time implementation.

## II. SBSS USING CTF FOR NAEC

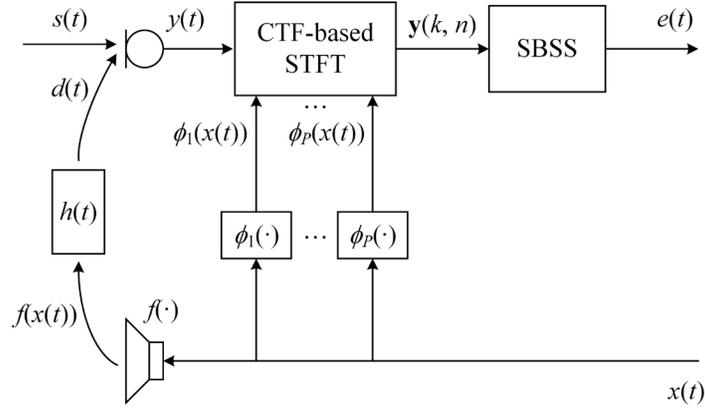

FIG. 1. SBSS using CTF for NAEC.

As depicted in Fig. 1, the microphone signal $y(t)$ with time index $t$ is expressed as

$$y(t) = d(t) + s(t) = h(t) * f(x(t)) + s(t), \quad (1)$$

where $x(t)$, $d(t)$, and $s(t)$ are the far-end signal, the nonlinear echo signal, and the near-end signal, respectively. $d(t)$ is generated by convolving the echo path $h(t)$ with the nonlinearly mapped far-end signal $f(x(t))$, which can be expressed using a basis-generic expansion model as [10]

$$f(x(t)) = \sum_{i=1}^{P} a_i \phi_i(x(t)), \tag{2}$$

where $\phi_i(\cdot)$ is the *i*-th order basis function weighted by the corresponding coefficient $a_i$ and $P$ is the expansion order. Here we use the odd power series [10, 11] for the nonlinear basis function as

$$\phi_i(x(t)) = x^{2i-1}(t), \tag{3}$$

since it can achieve better performance than other expansions with the same order, such as Fourier series and even power series.

Substituting Eq. (2) into Eq. (1) results in a time-domain observation model as

$$y(t) = h(t) * \left[ \sum_{i=1}^{P} a_i \phi_i(x(t)) \right] + s(t), \tag{4}$$

which can be further converted into the STFT representation using the MTF approximation as in Ref. 14. To reduce the latency caused by MTF, we use the CTF approximation [25, 26] instead, with the resulting STFT-domain observation model as

$$\begin{aligned} Y(k,n) &= \sum_{l=0}^{L-1} H_l(k,n) \left[ \sum_{i=1}^{P} a_i X_{\phi,i}(k,n-l) \right] + S(k,n) \\ &= \sum_{i=1}^{P} \sum_{l=0}^{L-1} H_{i,l}(k,n) X_{\phi,i}(k,n-l) + S(k,n), \end{aligned} \tag{5}$$

where $Y(k, n)$, $X_{\phi,i}(k, n)$, and $S(k, n)$ are the STFT-domain representations of $y(t)$, $\phi_i(x(t))$, and $s(t)$ with the frequency index $k$ and the frame index $n$, respectively. $H_{i,l}(k, n) = a_i H_l(k, n)$ is the mixed nonlinear acoustic transfer function and $L$ is the number of short-time frames.

By defining $L \times 1$ vectors $\mathbf{h}_i(k, n) = [H_{i,0}(k, n), \ldots, H_{i,L-1}(k, n)]^{\mathrm{T}}$ and $\mathbf{x}_i(k, n) = [X_{\phi,i}(k, n), \ldots, X_{\phi,i}(k, n - L + 1)]^{\mathrm{T}}$, where the superscript T denotes the transpose operation, Eq. (5) can be written as

$$Y(k,n) = \sum_{i=1}^{P} \mathbf{h}_i^{\mathrm{T}}(k,n) \mathbf{x}_i(k,n) + S(k,n). \tag{6}$$

By defining the mixed output vector $\mathbf{y}(k, n) = [Y(k, n), \mathbf{x}_1^T(k, n), \ldots, \mathbf{x}_P^T(k, n)]^T$ and the mixed source vector $\mathbf{s}(k, n) = [S(k, n), \mathbf{x}_1^T(k, n), \ldots, \mathbf{x}_P^T(k, n)]^T$, both with dimension $(PL + 1) \times 1$, Eq. (6) can be expressed in the vector-matrix form as

$$\mathbf{y}(k,n) = \mathbf{H}(k,n)\mathbf{s}(k,n), \tag{7}$$

with

$$\mathbf{H}(k,n) = \begin{bmatrix} 1 & \mathbf{h}^T(k,n) \\ \mathbf{0}_{PL \times 1} & \mathbf{I}_{PL} \end{bmatrix}, \tag{8}$$

where $\mathbf{H}(k, n)$ is the mixing matrix of size $(PL + 1) \times (PL + 1)$, $\mathbf{h}(k, n) = [\mathbf{h}_1^T(k, n), \ldots, \mathbf{h}_P^T(k, n)]^T$ is the mixing vector of size $PL \times 1$, $\mathbf{0}_{PL \times 1}$ is a zero vector of size $PL \times 1$, and $\mathbf{I}_{PL}$ is an identity matrix of size $PL \times PL$.

The target of the echo cancellation is to remove the echo and extract the near-end signal from the microphone signal. All the $\mathbf{x}_i(k, n)$ can be regarded as the known reference signals. From the viewpoint of BSS, only the microphone signal should be separated into the echo and the near-end signal, while the reference signals keep fixed. The estimate $E(k, n)$ of the near-end signal $S(k, n)$ can be obtained following the demixing process as

$$\mathbf{e}(k,n) = \mathbf{W}(k,n)\mathbf{y}(k,n), \tag{9}$$

with

$$\mathbf{W}(k,n) = \begin{bmatrix} 1 & \mathbf{w}^T(k,n) \\ \mathbf{0}_{PL \times 1} & \mathbf{I}_{PL} \end{bmatrix}, \tag{10}$$

where $\mathbf{e}(k, n) = [E(k, n), \mathbf{x}_1^T(k, n), \ldots, \mathbf{x}_P^T(k, n)]^T$ is the estimated vector of size $(PL + 1) \times 1$, $\mathbf{W}(k, n)$ is the demixing matrix of size $(PL + 1) \times (PL + 1)$, and $\mathbf{w}(k, n)$ is the demixing vector of size $PL \times 1$.

Note that although the reference signals $\mathbf{x}_i(k, n)$ are not independent, the near-end signal is still independent of all these reference signals, so that SBSS can be utilized to extract the near-end signal with the constrained demixing matrix [14].

## III. SBSS ALGORITHMS

Inspired by AuxIVA [18–21] and ILRMA [22–24], two effective methods commonly used in the BSS problems, which have been proven to have better performance than IVA using natural gradient [15, 16], we propose two SBSS algorithms based on their optimization strategies to optimize the demixing matrix.

### A. AuxIVA-based SBSS

AuxIVA has stable and fast update rules based on the auxiliary function technique [18]. The update rules of $\mathbf{W}(k, n)$ in Eq. (9) for the offline AuxIVA are expressed as follows [19]:

$$r_m(n) = \sqrt{\sum_{k=1}^{K} \left| \mathbf{w}_m^H(k) \mathbf{y}(k, n) \right|^2}, \tag{11}$$

$$\mathbf{V}_m(k) = \frac{1}{N} \sum_{n=1}^{N} \frac{G'(r_m(n))}{r_m(n)} \mathbf{y}(k, n) \mathbf{y}^H(k, n), \tag{12}$$

$$\mathbf{w}_m(k) \leftarrow \left( \mathbf{W}(k) \mathbf{V}_m(k) \right)^{-1} \mathbf{e}_m, \tag{13}$$

$$\mathbf{w}_m(k) \leftarrow \frac{\mathbf{w}_m(k)}{\sqrt{\mathbf{w}_m^H(k) \mathbf{V}_m(k) \mathbf{w}_m(k)}}, \tag{14}$$

where $\mathbf{w}_m^H(k)$ denotes the $m$-th row of the offline demixing matrix $\mathbf{W}(k) = [\mathbf{w}_1(k), \ldots, \mathbf{w}_{PL+1}(k)]^H$, the superscript H denotes the Hermitian transpose, $r_m(n)$ is the auxiliary variable, $K$ is the number of frequency bins, $\mathbf{V}_m(k)$ is the weighted covariance matrix, $N$ is the number of time frames, $G'(r_m(n))$ is the derivative of the contrast function [19], and $\mathbf{e}_m$ denotes the one-hot vector whose $m$-th element is 1.

It can be seen from the constrained structure of **W**(*k*, *n*) in Eq. (10) that only the first row of **W**(*k*, *n*) needs to be updated and the first element should be fixed as 1. Also, the online AuxIVA [21] is required for real-time NAEC. Thus, the update rules for the AuxIVA-based SBSS are derived as follows:

$$r_1(n) = \sqrt{\sum_{k=1}^{K} \left| \mathbf{w}_1^H(k,n) \mathbf{y}(k,n) \right|^2}, \tag{15}$$

$$\mathbf{V}_1(k,n) = \alpha \mathbf{V}_1(k,n-1) + (1-\alpha) \Phi(r_1(n)) \mathbf{y}(k,n) \mathbf{y}^H(k,n), \tag{16}$$

$$\mathbf{w}_1(k,n) \leftarrow \left( \mathbf{W}(k,n) \mathbf{V}_1(k,n) \right)^{-1} \mathbf{e}_1 = \mathbf{V}_1^{-1}(k,n) \mathbf{e}_1, \tag{17}$$

$$\mathbf{w}_1(k,n) \leftarrow \frac{\mathbf{w}_1(k,n)}{w_{1,1}(k,n)}, \tag{18}$$

where $\alpha$ is a forgetting factor, $\Phi(r_1(n)) = r_1^{\beta-2}(n)$ denotes a weighting function [20], $\beta$ is a shape parameter, and $w_{1,1}(k, n)$ denotes the first element of $\mathbf{w}_1(k, n)$.

### B. ILRMA-based SBSS

ILRMA can be regarded as a further refinement of IVA by utilizing nonnegative matrix factorization (NMF) [27–29] as a more flexible source model, which can capture the spectral structures that cannot be used by IVA. The update rules of the source model for ILRMA are expressed as follows [22–24]:

$$t_m(k,b) \leftarrow t_m(k,b) \sqrt{\frac{\sum_{n=1}^{N} \left| e_m(k,n) \right|^2 v_m(b,n) r_m^{-2}(k,n)}{\sum_{n=1}^{N} v_m(b,n) r_m^{-1}(k,n)}}, \tag{19}$$

$$v_m(b,n) \leftarrow v_m(b,n) \sqrt{\frac{\sum_{k=1}^{K} \left| e_m(k,n) \right|^2 t_m(k,b) r_m^{-2}(k,n)}{\sum_{k=1}^{K} t_m(k,b) r_m^{-1}(k,n)}}, \tag{20}$$

$$r_m(k,n) = \sum_{b=1}^{B} t_m(k,b) v_m(b,n), \tag{21}$$

where $t_m(k, b)$ and $v_m(b, n)$ are the source-wise bases and activations with the basis index $b$, respectively, $B$ is the number of bases, $e_m(k, n)$ denotes the $m$-th element of $\mathbf{e}(k, n)$, and $r_m(k, n)$ is the estimated source-wise variance, which should be updated by Eq. (21) after each update of $t_m(k, b)$ and $v_m(b, n)$.

The update rules of the demixing matrix for ILRMA are similar to those of AuxIVA as follows:

$$\mathbf{V}_m(k) = \frac{1}{N} \sum_{n=1}^{N} \frac{1}{r_m(k,n)} \mathbf{y}(k,n) \mathbf{y}^H(k,n), \tag{22}$$

$$\mathbf{w}_m(k) \leftarrow \left(\mathbf{W}(k) \mathbf{V}_m(k)\right)^{-1} \mathbf{e}_m, \tag{23}$$

$$\mathbf{w}_m(k) \leftarrow \frac{\mathbf{w}_m(k)}{\sqrt{\mathbf{w}_m^H(k) \mathbf{V}_m(k) \mathbf{w}_m(k)}}. \tag{24}$$

For the SBSS, the first row of $\mathbf{W}(k, n)$ can be updated in the same way as in the AuxIVA-based SBSS, considering the constrained structure of the demixing matrix. For the real-time implementation, the summation of all the time frames in Eq. (19) can be replaced by the calculation only on the current frame. Thus, the update rules for the ILRMA-based SBSS are derived as follows:

$$t_1(k,b) \leftarrow t_1(k,b) \sqrt{\frac{|e_1(k,n)|^2 v_1(b,n) r_1^{-2}(k,n)}{v_1(b,n) r_1^{-1}(k,n)}}, \tag{25}$$

$$v_1(b,n) \leftarrow v_1(b,n) \sqrt{\frac{\sum_{k=1}^{K} |e_1(k,n)|^2 t_1(k,b) r_1^{-2}(k,n)}{\sum_{k=1}^{K} t_1(k,b) r_1^{-1}(k,n)}}, \tag{26}$$

$$r_1(k,n) = \sum_{b=1}^{B} t_1(k,b) v_1(b,n), \tag{27}$$

$$\mathbf{V}_1(k,n) = \alpha \mathbf{V}_1(k,n-1) + (1-\alpha) \frac{1}{r_1(k,n)} \mathbf{y}(k,n) \mathbf{y}^H(k,n), \tag{28}$$

$$\mathbf{w}_1(k,n) \leftarrow \mathbf{V}_1^{-1}(k,n)\mathbf{e}_1, \qquad (29)$$

$$\mathbf{w}_1(k,n) \leftarrow \frac{\mathbf{w}_1(k,n)}{w_{1,1}(k,n)}. \qquad (30)$$

## IV. EXPERIMENTAL RESULTS

We test our proposed methods on both simulated data and real recordings. The proposed AuxIVA-based SBSS (SBSS-AuxIVA) and ILRMA-based SBSS (SBSS-ILRMA) algorithms are compared with the conventional SBSS algorithm based on the natural gradient IVA (SBSS-NGIVA) [14] and the single-microphone form of the NAEC algorithm proposed in Ref. 12 using state-space modeling (SSM-NAEC). SBSS-NGIVA uses both the original MTF model and the proposed CTF model. SSM-NAEC uses the CTF model. In all the experiments, the STFT is implemented using a Hanning window of 1024 taps with 75% overlap at the 16 kHz sampling frequency, resulting in the same latency for the MTF-based and CTF-based algorithms. For the nonlinear expansion order $P$, a larger value may achieve better performance but increase the computational complexity, and $P = 3$ is a good choice to balance the performance and the complexity for our experiments. The number of short-time frames $L$ is 1 for the MTF-based algorithm. For the CTF-based algorithms, a larger $L$ may achieve better performance for highly reverberant environments but increase the computational complexity, and $L$ is set to 3 to balance the performance and the complexity. For the other parameter settings, the forgetting factor $\alpha$ is set to 0.99 for both SBSS-AuxIVA and SBSS-ILRMA, the shape parameter $\beta$ is set to 0.4 for SBSS-AuxIVA, and the number of bases $B$ is set to 10 for SBSS-ILRMA. SBSS-AuxIVA and SBSS-ILRMA are both updated once per frame. Exemplary audio samples are available online at https://github.com/ChengGuoliang0/audio-samples2.

### A. Simulations

In the simulations, the hard clipping function [10,12] is used to generate the nonlinearly mapped far-end signal, which is expressed as

$$f(x(t)) = \begin{cases} -x_{max}, & x(t) < -x_{max} \\ x(t), & |x(t)| \leq x_{max} \\ x_{max}, & x(t) > x_{max} \end{cases} \quad (31)$$

where the clipping threshold is set to $x_{max} = 0.2\max|x(t)|$.

The echo path is simulated by a room impulse response using the image method [30], where the reverberation time $T_{60}$ varies from 0.2 s to 0.8 s with an increment of 0.1 s. A white Gaussian noise is used to represent the background noise with signal-to-noise ratio (SNR) of 60 dB. We consider both the single-talk and double-talk cases with speech signals of 10-s duration. A male speech signal is used as the far-end signal for both cases. A female speech signal is used as the near-end signal for the double-talk case with signal-to-echo ratio (SER) of 0 dB. For the single-talk case, the performance is evaluated by the echo return loss enhancement (ERLE), defined as $10\log_{10}\{E[y^2(t)]/E[e^2(t)]\}$ [10]. For the double-talk case, we measure the true ERLE (tERLE), defined as $10\log_{10}\{E[d^2(t)]/E[(e(t)-s(t))^2]\}$ [10].

Figures 2 and 3 show the performance of the three CTF-based SBSS algorithms and the MTF-based SBSS-NGIVA over various reverberant conditions for the single-talk and double-talk cases. It can be seen that the CTF-based SBSS-NGIVA consistently outperforms the MTF-based SBSS-NGIVA, especially in highly reverberant conditions. Moreover, with the CTF model, SBSS-AuxIVA and SBSS-ILRMA both show better performance than SBSS-NGIVA. Figures 4 and 5 display the performance of the three CTF-based SBSS and SSM-NAEC algorithms with $T_{60} = 0.3$ s. Besides tERLE, we also use the perceptual evaluation of speech quality (PESQ) [31] and the short-time objective intelligibility (STOI) [32] as the objective measures for the double-talk case, whose results are shown in Table I. The benefit of SBSS-AuxIVA and SBSS-ILRMA, especially in the double-talk case, can be clearly seen. The ERLE and tERLE performances of the four algorithms with SNR = 30 dB and $T_{60}$ = 0.3 s are shown in Figs. 6 and 7, respectively. The proposed SBSS-AuxIVA and SBSS-ILRMA also achieve better performance than the other two algorithms in the noisy condition. Besides speech, we also consider the double-talk music case with SER = 0 dB and $T_{60}$ = 0.3 s, where the far-end and near-end signals are both music with 10-s duration. The corresponding tERLE performance is shown in Fig. 8. SBSS-ILRMA significantly outperforms the other algorithms as NMF used in ILRMA can effectively capture the spectral structures of music signals.

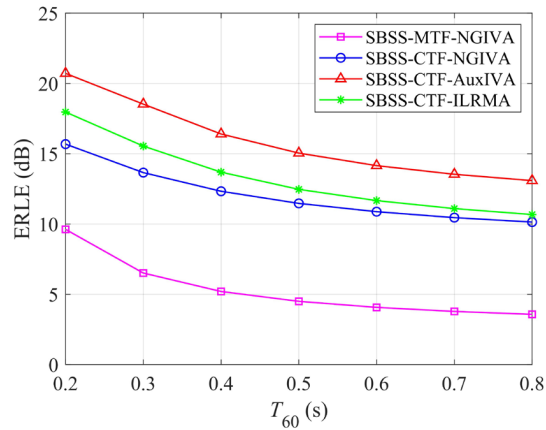

FIG. 2. (Color online) ERLE performance versus $T_{60}$ for the single-talk case.

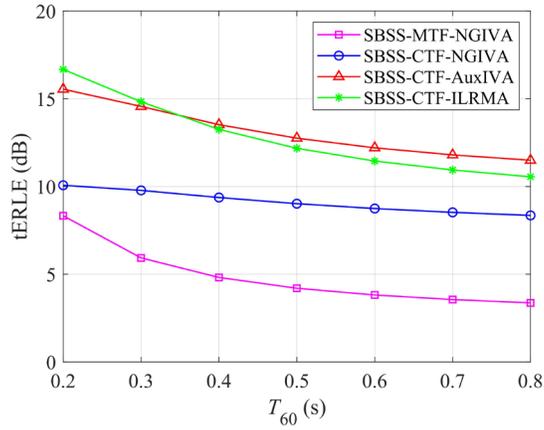

FIG. 3. (Color online) tERLE performance versus $T_{60}$ for the double-talk case.

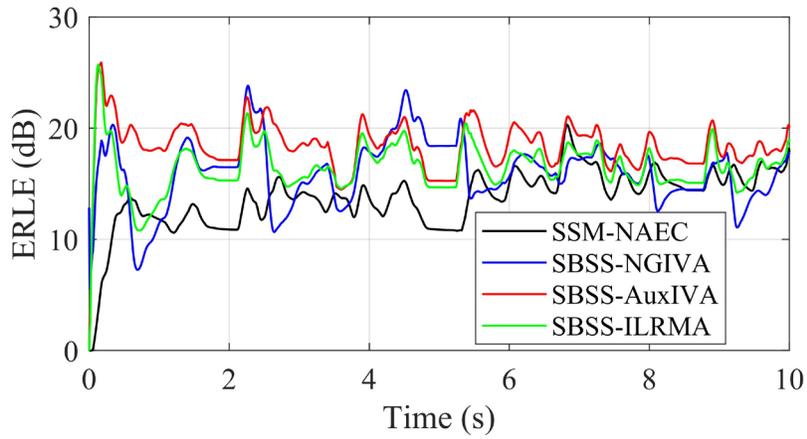

FIG. 4. (Color online) ERLE performance with $T_{60} = 0.3$ s for the single-talk case.

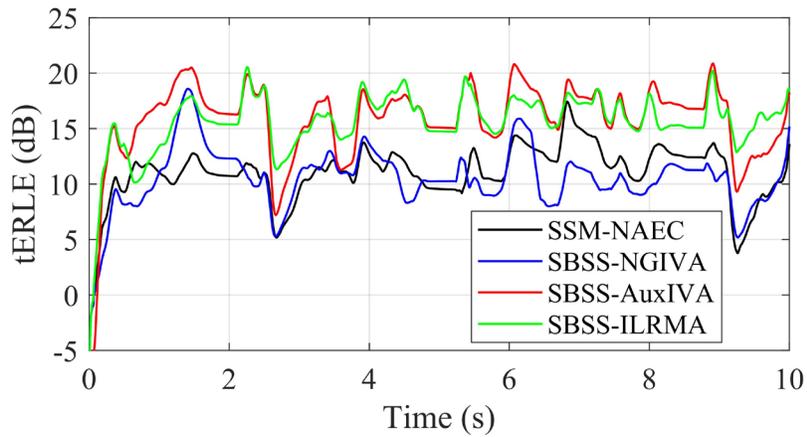

FIG. 5. (Color online) tERLE performance with $T_{60} = 0.3$ s for the double-talk case.

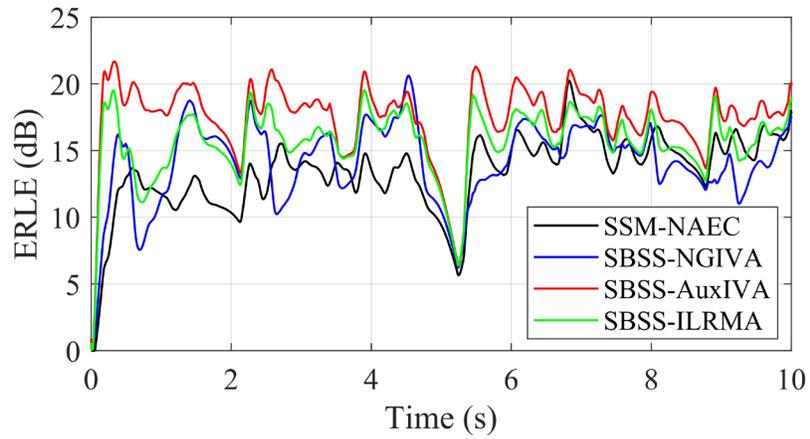

FIG. 6. (Color online) ERLE performance with SNR = 30 dB for the single-talk case.

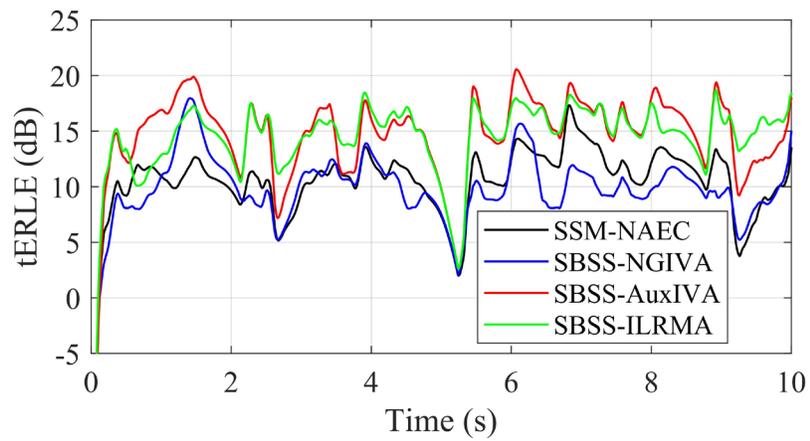

FIG. 7. (Color online) tERLE performance with SNR = 30 dB for the double-talk case.

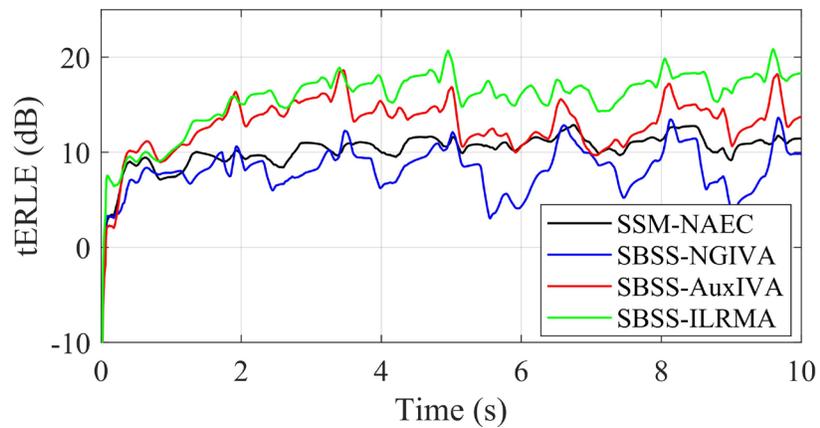

FIG. 8. (Color online) tERLE performance for the double-talk music case.

TABLE I. PESQ and STOI results for the double-talk case.

| Algorithms | PESQ | STOI |
|---|---|---|
| SSM-NAEC | 1.49 | 0.87 |
| SBSS-NGIVA | 1.44 | 0.85 |
| SBSS-AuxIVA | 1.73 | 0.92 |
| SBSS-ILRMA | 1.89 | 0.92 |

## B. Experiments on real recordings

We also evaluate the performance of the proposed methods using real recordings. The nonlinear echo signal is reproduced using a low-cost off-the-shelf small loudspeaker in an office with about 0.5 s reverberation time. The far-end signal is a 10-s long female speech signal, and the near-end signal is a 10-s long male speech signal for the double-talk case with SER = 0 dB. Figures 9 and 10 show the ERLE and tERLE performances for the single-talk and double-talk cases, respectively. The PESQ and STOI results for the double-talk case are shown in Table II. Similar to the simulation results, SBSS algorithms converges faster than SSM-NAEC in the single-talk case, and SBSS-AuxIVA and SBSS-ILRMA have significantly better steady performance in the double-talk case, both in terms of tERLE and the objective near-end speech quality metrics. Our MATLAB code can finish the test on 10 s audio sample in 5.6 s for SBSS-AuxIVA and 5.8 s for SBSS-ILRMA on a laptop with Intel Core i5-11320H CPU and 16 GB memory, validating the feasibility of real-time implementation of the proposed algorithms.

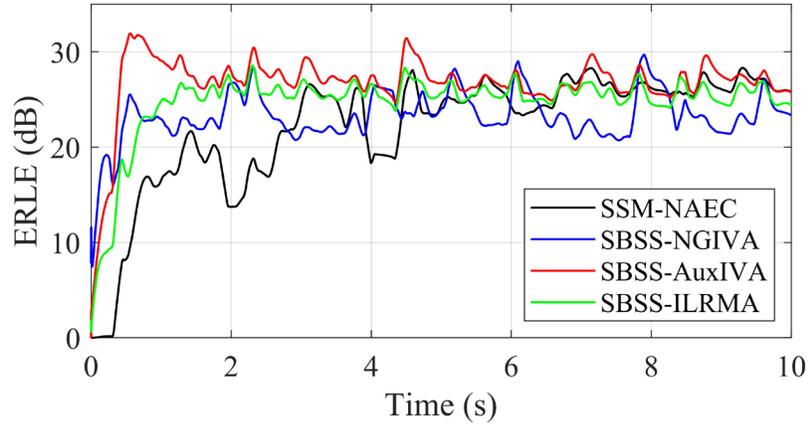

FIG. 9. (Color online) ERLE performance using recorded speech for the single-talk case.

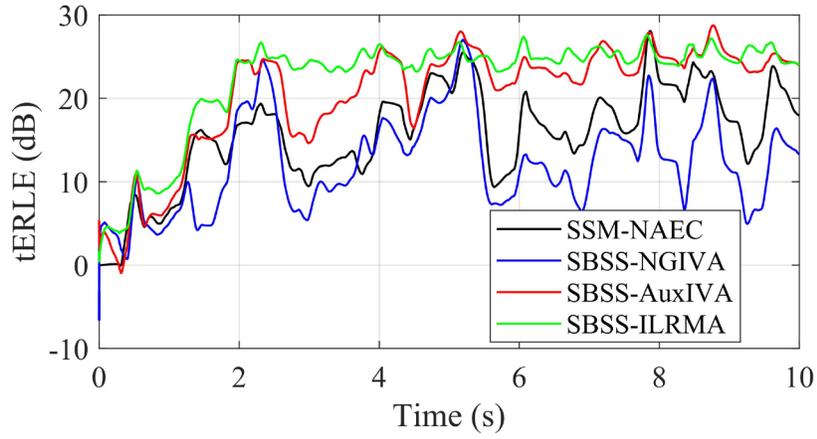

FIG. 10. (Color online) tERLE performance using recorded speech for the double-talk case.

TABLE II. PESQ and STOI results using recorded speech for the double-talk case.

| Algorithms | PESQ | STOI |
| --- | --- | --- |
| SSM-NAEC | 2.06 | 0.96 |
| SBSS-NGIVA | 1.95 | 0.91 |
| SBSS-AuxIVA | 2.61 | 0.97 |
| SBSS-ILRMA | 2.77 | 0.98 |

## V. CONCLUSION

In this paper, we propose two SBSS methods based on the optimization strategies of AuxIVA and ILRMA for NAEC using the CTF approximation. We use the CTF approximation to reduce the latency, making the SBSS more suitable for real-time NAEC applications. The update rules of the proposed methods are designed by carefully regularizing the AuxIVA and ILRMA algorithms according to the constrained structure of the demixing matrix. Experimental results on both simulated data and real recordings show that the proposed methods achieve better performance than the conventional SBSS method.

**ACKNOWLEGEMENTS**

This work was supported by the National Natural Science Foundation of China (Grant No. 11874219).

[32] C. H. Taal, R. C. Hendriks, R. Heusdens, and J. Jensen, "A short-time objective intelligibility measure for time-frequency weighted noisy speech," in *Proc. IEEE Int. Conf. Acoust., Speech, Signal Process.* (2010), Mar., Dallas, TX, USA, pp. 4214–4217.